\documentclass[twocolumn]{jpsj2} 
\title{The quantum critical point in CeRhIn$_5$: a resistivity study}

\author{Georg \textsc{Knebel}\thanks{E-mail: georg.knebel@cea.fr}, Dai \textsc{Aoki}, Jean-Pascal \textsc{Brison}, and Jacques \textsc{Flouquet}}

\inst{Commissariat \`a l' \'Energie Atomic, INAC, SPSMS, 17 rue des Martyrs, 38054 Grenoble, France}

\abst{The pressure--temperature phase diagram of CeRhIn$_5$ has been studied under high magnetic field by resistivity measurements. Clear signatures of a quantum critical point has been found at a critical pressure of $p_{\rm c}\approx 2.5$~GPa. The field induced magnetic state in the superconducting state is stable up to the highest field. At $p_{\rm c}$ the antiferromagnetic ground-state under high magnetic field collapses very rapidly. Clear signatures of $p_{\rm c}$ are the strong enhancement of the resistivity in the normal state and of the inelastic scattering term. No clear $T^2$ temperature dependence could be found for pressures above $T_c$. From the analysis of the upper critical field within a strong coupling model we present the pressure dependence of the coupling parameter $\lambda$ and the gyromagnetic ratio $g$. No signatures of a spatially modulated order parameter could be evidenced. A detailed comparison with the magnetic field--temperature phase diagram of CeCoIn$_5$ is given. The comparison between CeRhIn$_5$ and CeCoIn$_5$ points out the importance to take into account the field dependence of the effective mass in the calculation of the superconducting upper critical field $H_{\rm c2}$. It suggests also that when the magnetic critical field $H_{\rm M}(0)$ becomes lower than $H_{\rm c2} (0)$, the persistence of a superconducting pseudo-gap may stick the antiferromagnetism to $H_{\rm c2} (0)$.

}

\kword{CeRhIn$_5$, heavy fermion superconductor, quantum critical point, upper critical field}

\begin{document}
\maketitle

\section{Introduction}
The interplay of long range magnetic order and superconductivity is one of the central questions in the physics of heavy fermion systems. Usually small amounts of magnetic impurities lead to suppress the superconducting state in conventional superconductors, while in several heavy fermion compounds it is found that superconductivity (SC) appears just close to a quantum phase transition or can even coexist with magnetic order.\cite{Flouquet2005, Thalmeier2005} It is generally believed that quantum fluctuations are responsible for the attractive interaction to form Cooper pairs. Both scenarios, magnetic fluctuations close to a quantum critical point (QCP) where long range magnetic order is suppressed,\cite{Mathur1998} as well as density fluctuations due to a valence transition can lead to an attractive interaction to form Cooper pairs.\cite{Miyake2007} Close to such a quantum phase transition the normal state properties show strong deviations from the usual Fermi liquid behavior of a metal at low temperature, notably the resistivity deviates strongly from the $T^2$ temperature dependence and the specific heat divided by temperature $\gamma = C/T$ increases to low temperatures.\cite{Loehneysen2007}

The heavy fermion family Ce$M$In$_5$ ($M$ = Co, Rh, or Ir) offers an ideal opportunity to study the competition between antiferromagnetism (AF) and SC.\cite{Sarrao2007} While CeCoIn$_5$ and CeIrIn$_5$ are superconducting at ambient pressure and antiferromagnetism can be induced either by doping on the $M$-site or on the In site,\cite{Pagliuso2002, Pham2006, Bauer2008} CeRhIn$_5$ is antiferromagnetically ordered below $T_{\rm N} = 3.8$~K at ambient pressure. It orders in an incommensurate magnetic structure with an ordering vector $\vec{q} = (1/2, 1/2, 0.297)$. In zero magnetic field AF is suppressed rapidly for pressures $p>p_{\rm c}^\star = 1.95$~GPa and the ground state is a purely superconducting with most probably $d$-wave symmetry.\cite{Hegger2000, Mito2001, Knebel2004, Park2006, Knebel2006, Yashima2007} At this pressure $p_{\rm c}^\star$ the antiferromagnetic transition temperatures and the superconducting transition temperature coincides, $T_{\rm N} = T_{\rm c} \approx 2.2$~K. It shows up that when $T_{\rm c}>T_{\rm N}$ no long range magnetic ordering can appear as at least large parts of the Fermi surface are gapped due to the onset of SC. Therefore, at zero magnetic field the QCP in CeRhIn$_5$ is hidden by SC. Below $p_{\rm c}^\star$ $(T_{\rm N} > T_{\rm c})$ coexistence of antiferromagnetism and SC is reported for $p>1$~GPa and even at ambient pressure.\cite{Zapf2002, Chen2006, Paglione2007} However, the nature of this superconducting state below $p_{\rm c}^\star$ is still under debate.\cite{Knebel2006} 

For pressures above $p_{\rm c}^\star$ the application of a magnetic field $H \parallel ab$ plane as well as for $H\perp ab$ leads to a new phase inside the superconducting state \cite{Park2006, Knebel2006, Park2007} which has been detected by ac calorimetry. This new phase is most probably a re-entrance of the magnetic phase. It is very reminiscent to the high magnetic field phase in CeCoIn$_5$.\cite{Bianchi2003b} However, in difference to CeCoIn$_5$, the re-entrance field seems to persist also for fields higher than the upper critical field $H_{\rm c2}$, as has been observed first in resistivity measurements.\cite{Muramatsu2001}. The field induced phase is suspected to collapse at the critical pressure $p_{\rm c} \approx 2.5$~GPa. Interestingly the shape of the Fermi surface, as detected in de Haas van Alphen experiments changes abruptly close to $p_{\rm c}$ and the effective mass of the observed orbits increases strongly in the vicinity of $p_{\rm c}$ \cite{Shishido2005}. A detailed study of the electrical transport properties under high pressure at rather high temperatures has been published recently.\cite{Nakashima2007}

In this paper we will give a detailed study of the low temperature electrical resistivity of CeRhIn$_5$ under high pressure and high magnetic field $H\parallel ab$. The aim will be to study the magnetic QCP by applying magnetic fields $H>H_{\rm c2}$ to suppress SC. Furthermore a detailed study of the pressure dependence of the upper critical field will be given and a comparison to CeCoIn$_5$ is given. 

\section{Experimental details}
The sample used in these experiments was cut from the same single crystal used in our specific heat experiments under high pressure \cite{Knebel2004, Knebel2006}. The dimension of the sample is 0.16 $\times$ 0.09  $\times$ 0.05 mm$^3$. At ambient pressure the residual resistivity ratio $\rho (300{\rm K})/\rho (0 {\rm K})\approx 200$ indicates the high quality of the sample. The electrical resistivity was measured using a standard four point lock-in technique at 17~Hz. Electrical contacts to the sample have been realized by spot-welding 10 $\mu$m Au wires to the sample. A current of maximal 100 $\mu$A was used to measure the resistivity at low temperature. The temperature was measured with a calibrated Ge thermometer which is fixed on the mixing chamber of the dilution refrigerator in a field compensated region of the cryostat. The pressure cell has been thermalized to the mixing chamber using a Cu rod with 10 mm diameter. A magnetic field of maximal 16~T could be applied within the ab plane of the crystal perpendicular to the current direction. 

High pressure measurements have been performed in a diamond anvil pressure cell with argon as pressure medium. The pressure has been fixed at ambient temperature and determined by measuring the fluorescence of ruby before and after the experiment at liquid nitrogen temperature. The difference of these pressure determination was less than 0.15 GPa in each case. 

\section{Results}
\subsection{Resistivity in zero magnetic field}

\begin{figure}
\begin{center}
\includegraphics[width=0.9\hsize,clip]{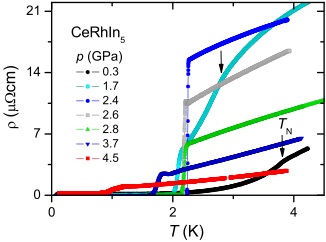}%
 \caption{\label{zeronearTc}(Color online) Resistivity of CeRhIn$_5$ in zero magnetic field for different pressures. The arrows indicate the magnetic transition for $p=0.3$ and 1.7~GPa.}
\end{center} 
 \end{figure}

Figure \ref{zeronearTc} shows the resistivity of CeRhIn$_5$ in zero magnetic field for different pressures. The antiferromagnetic transition for pressures below $p_{\rm c}^\star=1.95$~GPa is clearly visible. No SC is observed for $p=0.3$~GPa in this sample. For $p=1.7$~GPa a superconducting transition appears at $T^{\rm onset}_{\rm c}=2.12$~K below $T_{\rm N}=2.65$~K. It is remarkable that $T_{\rm c}$ determined from resistivity appears much higher in temperature  compared to the the previous specific heat experiment performed on a sample cut from the same single crystal with $T_{\rm c} (C) = 1.27$~K \cite{Knebel2006}. Such a discrepancy of the transition temperatures on an identical sample has been already observed in previous NQR experiment at $p=1.72$~GPa where the onset of $T_{\rm c}$ detected by the ac susceptibility at $T_{\rm c}^{\rm onset}= 2$~K but the mean field transition is at lower temperature $T_{\rm c}^{\rm MF} = 0.9$~K determined from NQR relaxation rate  \cite{Kawasaki2003}. Thus the observation that at least the appearance of superconductivity in the pressure range below $p_{\rm c}^\star$ is inhomogeneous seems to be a general feature.
Above 2~GPa, close to $p_{\rm c} \approx 2.5$~GPa, a very sharp superconducting transition is observed with a width of $\Delta T_{\rm c} \approx 30$~mK. At high pressure $p > p_{\rm c}$, the superconducting transition broadens remarkably and $T_{\rm c}$ decreases. No superconductivity is observed above $5$~GPa.

\subsection{Resistivity under magnetic field}

\begin{figure}
\begin{center} 
\includegraphics[width=0.8\hsize,clip]{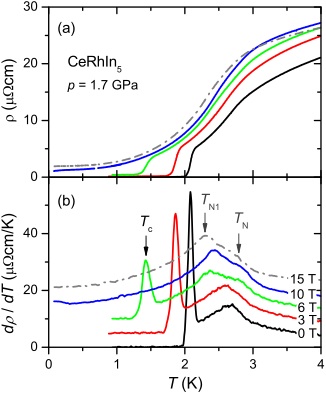}%
 \caption{\label{rho17kbar}(Color online) (a) Resistivity of CeRhIn$_5$ at $p=1.7$~GPa for different magnetic fields $H \perp c$. (b) Derivative $d\rho/dT$ of the resistivity versus temperature. The arrows indicate the temperature of the magnetic transitions or the superconducting transition. Curves are shifted by 5~$\mu\Omega$cm/K respectively.}
\end{center}
\end{figure}

\begin{figure}
\begin{center}
\includegraphics[width=0.8\hsize,clip]{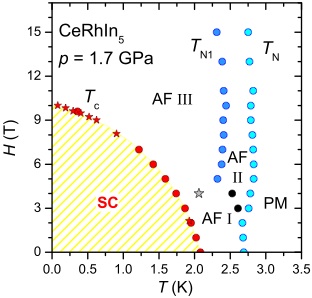}%
 \caption{\label{pd17kbar}(Color online) Phase diagram of CeRhIn$_5$ at $p=1.7$~GPa derived from the present resistivity measurements. Three different magnetic phases can be distinguished, the labeling corresponds to the different magnetic phases obtained at ambient pressure \cite{Cornelius2001, Raymond2007}. The superconducting transition has been derived from the midpoint of the transition. Stars corresponds to field sweeps, circles to temperature sweeps.}
 \end{center}
 \end{figure}

Next we will discuss the resistivity under magnetic field for a fixed pressure. Fig. \ref{rho17kbar}(a) shows the resistivity  at $p=1.7$~GPa as function of temperature. To determine the phase diagram, we plotted the derivative $d\rho / dT$ $vs.$ temperature in Fig.~\ref{rho17kbar}(b). At low fields $H<3$~T one magnetic transition appears at $T_{\rm N}$ above the superconducting transition at $T_{\rm c}$. For higher fields, two distinct magnetic anomalies can be seen in the derivative. From this data the phase diagram can be drawn, as shown in Fig.~\ref{pd17kbar}; it is reminiscent to the one obtained at ambient pressure \cite{Cornelius2001, Raymond2007} where three different magnetic phases can be distinguished. At zero pressure it has been shown in detailed neutron scattering experiments, that the incommensurate magnetic structure of phase AF I with an ordering vector $\vec{q}_{ic} = $(1/2, 1/2, 0.298) gets commensurable (phase AF III) under magnetic field at low temperatures with $\vec{q}_c = $(1/2, 1/2, 1/4) \cite{Raymond2007}. Phase AF II at ambient pressure has the same structure than the incommensurate phase AF I, but the ordered moment is reduced.  
The onset of superconductivity does not allow to draw the phase line between the antiferromagnetic phases AF I and AF III to lower temperatures. 
Remarkably, no accident can be observed in the $T$ dependence of the upper critical field $H_{\rm c2}$ close to the crossing point of the phase line of the incommensurate to commensurate transition (phase AF I to AF III) and the $H_{\rm c2} (T)$ line. It seems as if the superconducting phase is superimposed to the magnetic phase diagram without interplay; the same phenomenon will appear above $p_c^\star$.

\begin{figure}
\begin{center}
\includegraphics[width=0.8\hsize,clip]{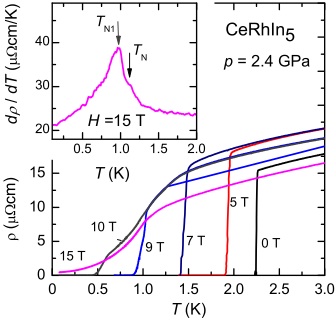}%
 \caption{\label{rho24kbar}(Color online) Resistivity of CeRhIn$_5$ at $p=2.4$~GPa for different magnetic fields. (Inset) Derivative $d\rho/dT$ of the resistivity versus temperature for $H=15$~T. Arrows indicate the temperature of the magnetic transitions at $T_{\rm N}$ and $T_{\rm N1}$.}
\end{center}
\end{figure}
 
\begin{figure}
\begin{center}
\includegraphics[width=0.8\hsize,clip]{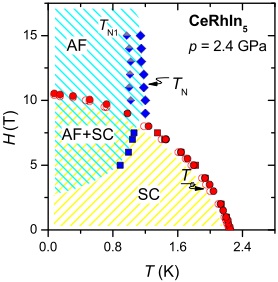}%
 \caption{\label{phase24kbar}(Color online) ($H, T$) phase diagram of CeRhIn$_5$ at $p=2.4$~GPa derived from the electrical resistivity (circles and diamonds) and from our previous ac calorimetry measurements (squares) \cite{Knebel2006}. The pressure of resistivity and specific heat measurement may be slightly different. Open and closed circles give zero resistivity and the midpoint of the superconducting transition in the resistivity; filled and half-filled diamonds correspond to $T_{\rm N}$ and $T_{\rm N1}$ determined from the derivative $d\rho / dT$, respectively. (See inset Fig.\ref{rho24kbar}.) }
\end{center}
\end{figure}

\begin{figure}
\begin{center}
\includegraphics[width=0.8\hsize,clip]{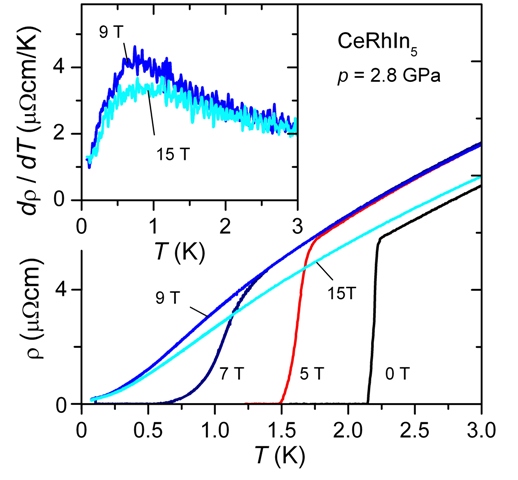}%
 \caption{\label{rho28kbar}(Color online)  Resistivity of CeRhIn$_5$ at $p=2.8$~GPa for different pressures. (Inset) Derivative $d\rho/dT$ of the resistivity versus temperature for $H=9$~T and 15~T. No magnetic anomaly can be seen in the resistivity measurements. }
\end{center}
\end{figure}

Increasing the pressure above $p_{\rm c}^\star$ leads to a superconducting ground state and in zero magnetic field the antiferromagnetism is suppressed. The main panel of Fig.~\ref{rho24kbar} presents the resistivity for different magnetic fields at $p=2.4$~GPa. The superconducting transition at low field is very sharp, broadening appears for fields above 7~T. The midpoint of the superconducting transition for 10~T is at $T_{\rm c} = 0.51$~K by a width of $\Delta T_{\rm c} \approx 140$~mK, $H_{\rm c2} (0)$ can be extrapolated to 10.62~T. For magnetic fields $H > 9$~T two further anomalies can be detected above the superconducting transition. The maximum of the derivative $d\rho / dT$ marks the transition temperature $T_{\rm N1}$ and the shoulder the N\'{e}el temperature $T_{\rm N}$ (see inset of Fig.~\ref{rho24kbar}). 
Even at the highest field, these two transitions can be observed. From these data together with previous specific heat results \cite{Knebel2006} the magnetic phase diagram at this pressure can be plotted as shown in Fig.~\ref{phase24kbar} (small differences in pressure between specific heat and resistivity data explain the small shift of $T_{\rm N}$ on crossing $H_{\rm c2}$). The application of a magnetic field leads to a phase transition inside the superconducting state where AF and SC coexist \cite{Knebel2006,Park2006}. However, the AF state is very stable, even far above $H_{\rm c2}$ and can be followed in the resistivity up to at least 15~T. Remarkably, again no anomaly in $H_{\rm c2} (T)$ occurs close to the crossing of $H_{\rm c2} (T)$ and $T_{\rm N} (H)$.

\begin{figure}
\begin{center}
\includegraphics[width=0.8\hsize,clip]{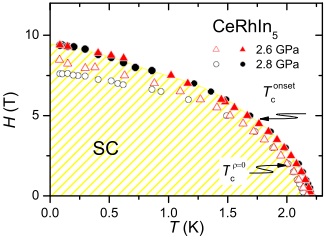}%
\caption{\label{phase28kbar}(Color online) Field--temperature phase diagram of CeRhIn$_5$ for $p=2.6$~GPa (triangles) and 2.8~GPa (circles) . Closed symbols mark the onset of the transition, open symbols correspond to the temperature of $\rho = 0$. With increasing pressure the width of the superconducting transition increases significantly to low temperatures.}
\end{center}
\end{figure}

Above the critical pressure $p_{\rm c} \approx 2.5$~GPa the antiferromagnetism is completely suppressed. Fig.~\ref{rho28kbar} presents the resistivity and the inset $d\rho / dT$ as function of temperature for $p=2.8$~GPa. No magnetic transition can be observed. The broad maximum in $d\rho / dT$ corresponds to the change of curvature in the resistivity and is not due to any magnetic anomaly. Even at zero field the superconducting transition is slightly broader than at the maximum of $T_{\rm c}$, $\Delta T_{\rm c} \approx 60$~mK; it is associated to the change in the slope of $dT_{\rm c} / dp$. With increasing magnetic field the transition broadens significantly.  In Fig.~\ref{phase28kbar} we indicate the onset of the transition and zero resistivity as function of magnetic field for $p=2.6$~GPa and 2.8~GPa. The pressure dependence of the upper critical field will be discussed below in detail.

\section{Discussion} 

\subsection{Pressure and field dependence in CeRhIn$_5$}

\begin{figure}
\begin{center}
\includegraphics[width=0.8\hsize,clip]{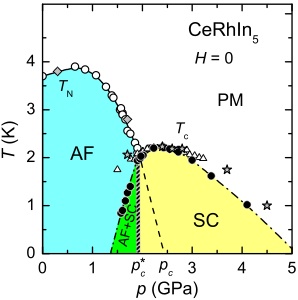}%
\caption{\label{phase_diagram}(Color online) Pressure--temperature phase diagram of CeRhIn$_5$ in zero magnetic field from ac calorimetry (circles) \cite{Knebel2006}, ac susceptibility (triangles)\cite{Knebel2004}  and resistivity (this work, diamonds for $T_{\rm N}$ and stars for $T_{\rm c}$). At low pressure the ground state is antiferromagnetic. Below $p_{\rm c}^\star$ both, antiferromagnetism (AF) and superconductivity (SC) coexists. At $p_{\rm c}^\star$ the AF is suppressed suddenly before the quantum critical point at $p_{\rm c}$ is reached under pressure. Above $p_{\rm c}^\star$ a purely superconducting the ground state appears in zero magnetic field. The dashed line gives the expected pressure dependence of the N\'{e}el temperature in absence of superconductivity. }
\end{center}
\end{figure}

The ($p, T, H$) phase diagram of CeRhIn$_5$ is extremely rich (see Fig. \ref{phase_diagram}).  At $p_{\rm c}^\star$, the two critical temperatures $T_{\rm N}$ and $T_{\rm c}$ merge into one point. In a first approach the crossing point looks like a bi-critical point: as function of pressure a direct transition from AF  to  SC 
occurs. In the phase diagram in Fig.~\ref{phase_diagram} such direct transition corresponds to the vertical hatched area, without the emergence of a AF+SC regime. However, in a real experiment such a transition is difficult to realize under pressure, inhomogeneities (in the pressure as well as in the sample) may always impede such a `clear' phase diagram. Due to inhomogeneities an AF+SC regime can appear; however, it would not be homogeneous and phase separation into AF and SC parts is expected. Another possibility is that $p_{\rm c}^\star$ is a tetracritical point.\cite{Kivelson2001, Demler2004} Strong support for this scenario comes from the homogeneous character of the nuclear spin dynamics in the AF+SC domain at low temperature ($T < T_{\rm c}$).\cite{Yashima2007} 
In recent nuclear-quadrupole-resonance (NQR) experiments the observation of a tetra-critical point in zero magnetic field has been reported and it has been suggested that a uniformly homogeneous AF+SC phase exist below $p_{\rm c}^\star$. The uniformly coexistence of AF and SC in this pressure range is also followed from the fact that the NQR relaxation $(1/T_1)$ is mono-exponential, independent on the investigated In-site.\cite{Yashima2007} This led to the suggestion that both, the antiferromagnetic and the superconducting order parameter are strongly coupled as it is proposed in the SO(5) theory.\cite{Zhang1997, Demler2004} However, the superconducting phase transition at $T_{\rm c}$ is at least inhomogeneous below $p_{\rm c}^\star$, as with different experimental probes different transition temperatures are detected. The vertical hatched line describes then only the trend that the AF+SC domain is highly non-symmetrical by respect to $p_{\rm c}^\star$: AF needs to disappear just above $p_{\rm c}^\star$. From experimental point of view it is very difficult to draw precisely the AF+SC boundary.

For $p<p_{\rm c}^\star \approx 2$~GPa the ground state has an antiferromagnetic component. The superconducting phase transition observed below $p_{\rm c}^\star$ in the resistivity is not bulk in nature. However zero resistivity has been observed at $p=1.7$~GPa and the upper critical field determined by the resistivity is rather large. The magnetic ordered state seems not to change dramatically under high pressure. The magnetic ($H$--$T$) phase diagram observed at $p=1.7$~GPa is qualitatively unchanged in comparison to low pressure with the appearance of different magnetic phases (see Fig.~\ref{pd17kbar}).  

At zero pressure it has been shown in detailed neutron scattering experiments, that the incommensurate magnetic structure of phase AF I with an ordering vector $\vec{q}_{ic} = $(1/2, 1/2, 0.298) gets commensurable (phase AF III) under magnetic field at low temperatures with $\vec{q}_c = $(1/2, 1/2, 1/4) \cite{Raymond2007}. Phase II at ambient pressure has the same structure than the incommensurate phase AF I, but the ordered moment is reduced.  To identify the magnetic structures under high pressure, neutron scattering or NMR experiments are indispensable. However, up to now no  successful neutron scattering experiments have been performed under application of magnetic field and high pressure for CeRhIn$_5$. No definite conclusion can be given on the magnetic ordering vector under pressure in the different phases. All neutron scattering experiments performed up to now report an incommensurate ordering vector up to 1.7~GPa in zero magnetic field \cite{Majumdar2002, Llobet2004, Raymond2008}. In the most recent neutron scattering experiments at 1.7~GPa $\vec{q}_{\rm ic} (1.7\;{\rm GPa}) = (1/2, 1/2, 0.4)$ has been observed in zero field at $T=0.4$~K inside the superconducting state.\cite{Raymond2008}
Nevertheless, from our transport measurements here and also from the ac calorimetry under pressure \cite{Knebel2006} the different magnetic phases seem almost unchanged under high pressure up to $p_{\rm c}^\star$.\cite{note} The onset of superconductivity does not allow to draw the phase line between the antiferromagnetic phases AF I and AF III to lower temperatures.

Up to now it is unclear, if the magnetic ordering changes its commensurability under pressure.  The observation of the magnetic signal by neutron scattering, however, is not a direct prove of coexistence of AF and SC on a microscopic scale as the detected magnetic intensity is an average of the magnetic moment in the crystal volume. More detailed microscopic informations can be obtained from NQR measurements. As discussed above, these experiments point to an uniform coexistence of both AF and SC in this pressure range.\cite{Yashima2007}. Furthermore, it is stated from measurements of the NQR spectra at the In(2) site that the magnetic structure in this coexistence regime should be commensurate \cite{Yashima2007b}. Regarding to the phase diagram in Fig.~\ref{pd17kbar} it is difficult to imagine that there is a profound change in the magnetic structure. However, for the doped systems CeRh$_{1-x}$Ir$_x$In$_5$ \cite{Yokoyama2006, Ohira2007} and CeRh$_{1-x}$Co$_x$In$_5$ \cite{Christianson2005} as well as for Cd doped CeCo(In$_{1-x}$Cd$_x$)$_5$ \cite{Nicklas2007} remarkably AF and SC coexists only when an AF ordering with commensurate $\vec{q}_{\rm AF} = (1/2, 1/2, 1/2)$ is observed which is the ordering vector of the cubic CeIn$_3$ \cite{Lawrence1980, Benoit1980}. This is also the characteristic wavevector  in CeCoIn$_5$ where a sharp spin resonance develops in the superconducting state.\cite{Stock2008} In doped systems, the commensurate ordering vector $\vec{q}_{c}$ of phase III of CeRhIn$_5$ has never been reported. 

\begin{figure}[bt]
\begin{center}
\includegraphics[width=0.8\hsize,clip]{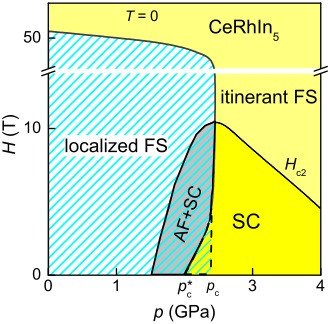}%
\caption{\label{fermi_surface}(Color online) $(H, p)$ phase diagram of CeRhIn$_5$ at $T \to 0$ indicating the Fermi surface topology in the different states of the phase diagram. The boundary between the localized Fermi surface (localized description of the 4$f$ electron) and of the itinerant paramagnetic phase (itinerant description of the 4$f$ electron) is indicated. One yet unsolved question is the Fermi surface topology in the AF+SC state with the strong interplay between antiferromagnetism and superconductivity. One can speculate that at $H = 0$ an itinerant FS persists down to $p_{\rm c}^\star$.  }
\end{center}
 \end{figure}

In the pressure range $p_{\rm c}^\star < p < p_{\rm c}$ at zero magnetic field,  the superconducting phase transition is well defined; bulk superconductivity appears at $p_c^\star$ and the antiferromagnetic state is rapidly suppressed. A natural explanation is that the opening of an superconducting gap on large parts of the Fermi surface leads to prevent the onset of long range antiferromagnetism. Spectacularly, the antiferromagnetic order is recovered inside the superconducting state under application of a magnetic field \cite{Park2006, Knebel2006}. In difference to $p<p_{\rm c}^\star$ in this pressure range the antiferromagnetic transition $T_{\rm N}$ is lower than the superconducting transition $T_{\rm c}$. The present resistivity measurements at $p=2.4$~GPa indicate that the antiferromagnetic state is robust up to high magnetic fields (see Fig.~\ref{phase24kbar}) and the field dependence of the antiferromagnetic transition $T_{\rm N}(H)$ and $T_{\rm c}(H)$ intersect in one point $(T^\star,H^\star)$. The re-entrant phase occurs for fields $H<H^\star$ as in the mixed state antiferromagnetism can be induced in the vortex core. A description of a homogeneous mixed superconducting and antiferromagnetic order parameter can be found in the frame of SO(5) theory \cite{Zhang1997, Demler2004}. The interesting effect is that antiferromagnetism can extend far from the vortex core. It is predicted that the antiferromagnetic signal will increase under magnetic field as the vortex number will be proportional to $H$. This results seems to be in agreement with the data of ref.~\citen{Park2006}. However, in our previous experiment the re-entrant signal disappears below at least 3~T.\cite{Knebel2006} The ac calorimetry experiments have evidenced clearly that the intersection point $(T^\star,H^\star)$ shifts to higher fields and lower temperatures as function of pressure \cite{Park2006, Knebel2006} and it was expected that the antiferromagnetically ordered phase collapses at the critical pressure $p_{\rm c} \approx 2.5$~GPa. 

For $p> p_{\rm c}$ indeed, no indication of re-entrance of antiferromagnetism under field is observed. The collapse of the antiferromagnetic state coincides with the strong change of the Fermi surface. (Small differences in the absolute value of the critical pressure $p_{\rm c}$ have been reported in various experiments, see e.g.~refs.~\citen{Shishido2005, Park2006, Knebel2006, Yashima2007}). 

A schematic ($H, p)$ phase diagram for $T = 0$ is shown in Fig.~\ref{fermi_surface} indicating the evolution of the Fermi surface under pressure and magnetic field. Up to the critical pressure ($p_{\rm c} \approx 2.35$~GPa in ref.~\citen{Shishido2005}) the Fermi surface of CeRhIn$_5$ is almost identical to that of the non-$4f$ reference compound LaRhIn$_5$ while the corresponding cyclotron masses increase strongly above $p=1.6$~GPa up to $p_{\rm c}$.\cite{Shishido2005} Thus the $4f$ electrons seems to be localized at the Ce site for $p < p_{\rm c}$. A distinct change of the de Haas van Alphen (dHvA) frequencies is observed for $p > p_{\rm c}$ and the Fermi surface of CeRhIn$_5$ is in good agreement with a $4f$-itinerant picture as in CeCoIn$_5$. This strong change of the Fermi surface seems to be connected to the rapid disappearance of the magnetism under magnetic field at $p_{\rm c}$. The dHvA oscillations are observed under high magnetic fields in the AF phase for $p < p_{\rm c}$ and in the paramagnetic phase $p > p_{\rm c}$.

However, even in zero magnetic field, the Fermi surface in the different phases AF, AF+SC, and inside the superconducting domain is still under debate;\cite{Miyake2006} above $p_{\rm c}^\star$ as $T_{\rm c}$ has a smooth pressure dependence without any anomaly at $p_{\rm c}$ it seems reasonable that the Fermi surface is already that of the paramagnetic phase with delocalized $f$ electrons as observed under high field above $p_{\rm c}$, see Fig.~\ref{fermi_surface}; in the antiferromagnetic domain below $p_{\rm c}^\star$, even when superconductivity occurs at $T_{\rm c} < T_{\rm N}$, the Fermi surface seems to be that detected in the pure antiferromagnetic state at high magnetic field.\cite{Shishido2005} 
However, in the field induced antiferromagnetic state for $p_{\rm c}^\star < p < p_{\rm c}$, the situation is not obvious at all. As $H_{\rm c2}(T)$ does not present any anomaly at the crossing point between the $H_{\rm c2}$ curve and the $T_{\rm N} (H)$ line (see Fig.\ref{phase24kbar}) one may guess that superconductivity continues to be developed on a Fermi surface characteristic of a paramagnetic phase. 
The Fermi surface in the AF+SC phase detected by dHvA measurements for $H\parallel c$ is still unclear.\cite{Ida2008} Thus one can speculate that the change of the Fermi surface in zero magnetic field (mainly $H \parallel c$) would occur not at $p_{\rm c}$ but even at the lower pressure $p_{\rm c}^\star$ where the antiferromagnetic state is rapidly suppressed in zero field, as previously discussed in ref.~\citen{Park2006}. 

\begin{figure}
\begin{center}
\includegraphics[width=0.8\hsize,clip]{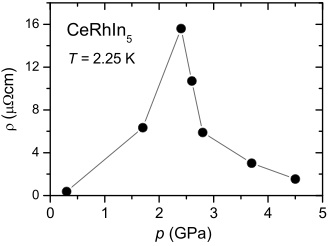}%
 \caption{\label{rho2,25K} Resistivity at $T=2.25$~K of CeRhIn$_5$ in zero magnetic field for different pressures. Close to the QCP at $p_{\rm c}\approx 2.5$~GPa the scattering is strongly enhanced.}
\end{center}
\end{figure}

In Fig.~\ref{rho2,25K} the pressure dependence of the resistivity at $T=2.25$~K is shown. A huge increase of the resistivity is observed near $p_{\rm c}$. This is a clear indication of a strong enhancement of the critical fluctuations due to an underlying quantum critical point \cite{Miyake2002}. Looking at the temperature dependence of the resistivity it can be clearly seen in Figs.~\ref{rho24kbar} and \ref{rho28kbar} that the temperature dependence is less than linear just above the transition temperature $T_{\rm c}$, as has been already shown in ref.~\citen{Knebel2007} for $p=2.7$~GPa as well as in ref.~\citen{Nakashima2007} and ref.\citen{Park2008}. However, one has to keep in mind that the temperature range $T>2$~K corresponds only to an cross-over regime. To study the normal state properties of CeRhIn$_5$ close to the critical pressure at low temperature high magnetic field is needed to suppress the superconducting state.  

\begin{figure}
\begin{center}
\includegraphics[width=0.8\hsize,clip]{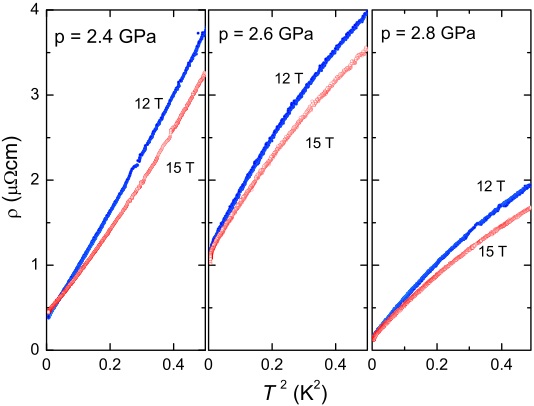}%
\caption{\label{rho_T2}(Color online) Resistivity of CeRhIn$_5$ at different pressures for a field of 12~T and 15~T as function of $T^2$ below $T=0.7$~K. 
For $p<p_{\rm c}\approx 2.5$~GPa the temperature }
\end{center}
\end{figure}

\begin{figure}
\begin{center}
\includegraphics[width=0.8\hsize,clip]{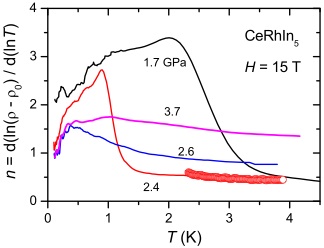}%
 \caption{\label{nT}(Color online) Temperature dependence of the resistivity exponent $n = d(\ln (\rho - \rho_0))/d\ln T$ for different pressures at high magnetic field (lines). Additionally we plotted $n(T)$ for $H=0$ at 2.4 GPa above the superconducting transition (circles). The field dependence of the exponent $n$ is rather weak. }
\end{center}
\end{figure}

Let us now look in more detail at the temperature variation of the resistivity. 
In Fig.~\ref{rho_T2} we have plotted the resistivity measured at a magnetic field of 12 and 15~T as function of $T^2$ below $T=0.7$~K. No ``clear'' Fermi liquid $T^2$ dependence is observed in this low temperature range. For $p =2.4$~GPa the curvature is positive indicating the presence of a magnon scattering term while it is negative for $p=2.6$~GPa. Also in this representation it gets very clear that the critical pressure is located between 2.4~GPa and 2.6~GPa.  
To analyze in more detail the temperature dependence of $\rho$ we calculated the temperature dependence of the exponent $n$ of a power law $\rho = \rho_0 + A_n T^n$ by $n = d(\ln (\rho - \rho_0))/d\ln T$. The result of this analysis is shown in Fig. \ref{nT} for different pressures at the highest measured fields $H = 15$~T. In the analysis an average of 50 data points (which corresponds to a temperature window of $T\approx 30$~mK at low temperature) is taken. To calculate the resistivity exponent at low magnetic fields in the temperature range above the superconducting transition in the normal state correctly, one has to know the value of the residual resistivity in absence of superconductivity. Taking the value of $\rho_0$ at high magnetic fields $H > H_{\rm c2}$ gives  $n \approx 0.4$ for $p=2.4$~GPa above 1.5~K for all fields. (see Fig.\ref{nT}). Thus $\rho(T)$ is only slightly field dependent in the normal state. 
At 1.7~GPa and 2.4~GPa the strong increase of $n$ at 15~T on cooling to a value $n>2$ indicates the onset of the antiferromagnetically ordered state. For $p>p_{\rm c}\approx 2.5$~GPa we find the exponent $n<2$ in all temperature range. At very low temperature, $n \to 1$ is found below 300~mK. Even away from the critical pressure at lowest temperatures the $T$ dependence is less than $T^2$. The difficulty to observe a nice $T^2$ dependence at very high pressure in the resistivity has already  been mentioned in ref.~\citen{Muramatsu2001}. 
We should emphasize that this not achievement of a $T^2$ dependence under high magnetic field at low temperature may 
result from a cross-over from a collision regime to a collision-less mode; the product $\omega_{\rm c} \tau$ of the cyclotron frequency $\omega_{\rm c}$ by the relaxation time $\tau$ becoming higher than 1 \cite{Taillefer1988a, Taillefer1988b}. In this limit, the characteristic time of the quasiparticle orbital motion is short compared with the time between
collisions, and therefore the Fermi surface topology plays an important role in determining transport properties. 

A linear temperature dependence of the resistivity close to a magnetic instability is generally taken as an indication of the importance of the quasi-two dimensional fluctuations in spin fluctuation theory (see e.g. ref.~\citen{Moriya2003}). The importance of a reduced dimensionality on the superconducting pairing strength has been discussed in refs.~\citen{Monthoux2001, Monthoux2004}. Of course, we should mention that  a linear $T$ dependent resistivity is also observed in a model of a critical valence transitions.\cite{Miyake2007, Onishi2000, Holmes2007} As discussed above, in CeRhIn$_5$ a strong change in the Fermi surface due to a delocalization of the 4$f$ electron is experimentally observed at $p_{\rm c}$ in full agreement with such a valence transition scenario which can also account for the enhancement residual resistivity at $p_{\rm c} \approx p_{\rm v}$ \cite{Miyake2002, Miyake2002b} and the appearance of superconductivity in a large range of pressure due to density fluctuations. 

\begin{figure}
\begin{center}
\includegraphics[width=0.8\hsize,clip]{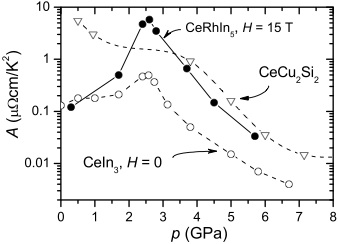}%
 \caption{\label{Acoeff_P} Pressure dependence of the $A$ coefficient of CeRhIn$_5$ at 15~T (full circles) compared to the pressure dependence of $A$  in CeIn$_3$ at zero field (open circles, from ref.~\citen{Knebel2001}) and in CeCu$_2$Si$_2$ (open triangles, from ref.~\citen{Vargoz1998}).} 
\end{center}
\end{figure}

\begin{figure}
\begin{center}
\includegraphics[width=0.8\hsize,clip]{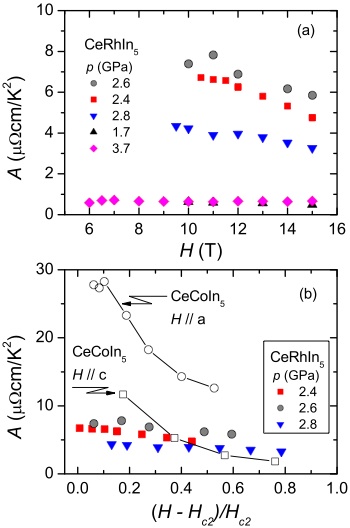}%
 \caption{\label{Afield}(Color online) (a) Field dependence of the $A$ coefficient of the resistivity in the normal state determined from a fit $\rho = \rho_0 + A T^2$ for CeRhIn$_5$ at different pressures for fields $H>H_{\rm c2}$ in the temperature range 0.1~K $< T <$ 0.4~K. (b) $A(H)$ as function of normalized field $(H - H_{\rm c2})/H_{\rm c2}$ of CeRhIn$_5$ close to the critical pressure $p_{\rm c}$ in comparison to CeCoIn$_5$ for $H \parallel a$ (open circles) and $H \parallel c$ (open squares). (Data for CeCoIn$_5$ at $p=0$~from ref.~\citen{Ronning2005}). Contrary to CeCoIn$_5$ \cite{Paglione2003, Ronning2005, Ronning2006}, no strong increase of the $A$ coefficient appears when approaching the upper critical field $H_{\rm c2}$.} 
\end{center}
\end{figure}

To analyze the field and pressure dependence of the quasi-particle scattering term $A$, we forced the resistivity data to be fitted by a  $A T^2$ dependence in the temperature range 0.1~K $< T <$ 0.4~K. The pressure dependence of $A$ is given in Fig.~\ref{Acoeff_P} for a field of $H=15$~T compared to $A$ in CeIn$_3$ and CeCu$_2$Si$_2$ at zero magnetic field.\cite{Knebel2001, Vargoz1998} For CeRhIn$_5$ the application of a high magnetic field is necessary to suppress the superconducting state. As discussed above in the normal state above the superconducting transition only a cross-over regime may be established. We plot the data for high field to be sure to be not to close to the superconducting transition. Clearly a pronounced maximum of the $A$ coefficient appears as function of pressure. At a low pressure of 0.3~GPa we find $A=0.12$~$\mu\Omega$cm/K$^2$ in CeRhIn$_5$ what is comparable to the value of CeIn$_3$ at zero pressure. This low pressure value of $A$ together with a Sommerfeld coefficient of the specific heat $\gamma = C/T \approx 56$~mJmol$^{-1}$K$^{-2}$ at ambient pressure gives a ratio $A / \gamma ^2 \approx 3.8 \times 10^{-5}$~$\mu\Omega$cm (mol K/mJ)$^2$ not far from the empirical universal value for heavy-fermion systems given by Kadowaki and Woods, $A / \gamma ^2 = 10^{-5}$~$\mu\Omega$cm (mol K/mJ)$^2$.\cite{Kadowaki1986} 
At ambient pressure CeRhIn$_5$ and CeIn$_3$ are antiferromagnetically ordered as the molecular field is very strong and the mass renormalization is rather weak. Under pressure the increase of $A$ is much stronger in CeRhIn$_5$ than in CeIn$_3$, $A=5.78$~$\mu\Omega$cm/K$^2$ at $p_{\rm c}$ for CeRhIn$_5$ while it increases only to $A=0.49$~$\mu\Omega$cm/K$^2$ for CeIn$_3$. In both compounds $A$ decreases significantly with pressure for $p > p_{\rm c}$. For CeRhIn$_5$ we find at the highest pressure of 5.6 GPa $A = 0.0335$~$\mu\Omega$cm/K$^2$. Contrary to these systems, CeCu$_2$Si$_2$ is located just at the magnetic quantum critical point at ambient pressure, thus the $A$ coefficient has its maximum at $p=0$. Clearly, in CeCu$_2$Si$_2$ a second critical pressure $p_{\rm v} \approx 4$~GPa indicates the critical valence fluctuations which are responsible for a second superconducting dome around $p_{\rm v}$ \cite{Holmes2004, Yuan2004}. The very small values of $A$ above $p_{\rm v}$ are a strong indication for a valence transition in Ce heavy fermion systems due to the strong increase of the hybridization with pressure and the concomitant loss of the crystal field degeneracy \cite{Tsujii2005}. In difference to CeCu$_2$Si$_2$ in CeRhIn$_5$ and CeIn$_3$ the critical pressures fall together, $p_{\rm c} \approx p_{\rm v}$.

The field dependence of the $A$ coefficient is plotted in Fig.~\ref{Afield}(a)  for different pressures. $A$ does not show a pronounced field dependence and this is also comparable to the observation in CeIn$_3$ \cite{Knebel2001}. Far from the critical pressure ($p=1.7$~GPa $<p_{\rm c}$ and $p=3.7$~GPa $>p_{\rm c}$) the field dependence of $A (H)$ is very flat. Close to the critical pressure $p_{\rm c} \approx 2.5$~GPa $A$ decreases almost linearly with increasing field $H > H_{\rm c2}$; however, there is no strong indication of quantum critical phenomena under magnetic field in this pressure range, in difference to CeCoIn$_5$ where a strong enhancement of $A$ close to $H_{\rm c2}$ for both field directions is observed if the current is perpendicular to the field direction as will be discussed below. 
Remarkably, similar field dependence of the effective mass are known in the dHvA experiments for both, CeRhIn$_5$ and CeCoIn$_5$.\cite{Settai2007}
Different magnetic field scales are important in different pressure domains of the phase diagram. In the antiferromagnetic domain $p<p_{\rm c}$ the relevant magnetic field scale is the critical field $H_M$ between the antiferromagnetic and the polarized paramagnetic domain under high field which is rather high ($H_{\rm M} = 52$~T at zero pressure \cite{Takeuchi2001} and $H_{\rm M} \gg 15$~T for $p = 2.4$~GPa). In the paramagnetic regime $p>p_{\rm c}$ it would be the field $H_{\rm K}$ which corresponds to the Kondo temperature ($\mu_{\rm B} H_{\rm K} \sim k_{\rm B} T_{\rm K}$). Of course for the superconducting properties the relevant scale is always the upper critical field. However, a coupling of these different scales is not compulsory. 

Insights of the superconducting properties can be  obtained by the analysis of the upper critical field $H_{\rm c2}$. Generally, the upper critical field is determined by the orbital and the paramagnetic pair-breaking effects. The orbital limiting field $H_{\rm orb}(T) = \Phi_0/2\pi \xi^2(T)$ is given by the fields at which vortex cores starts to overlap ($\Phi_0$ is the flux quantum). Close to $T_c$, it is always the dominant mechanism (the paramagnetic limitation has infinite slope at $T_c$), so that the initial slope of $H_{\rm c2}$ at $T_c$ is a good measure of the average Fermi velocity in directions perpendicular to the applied field : $H_{\rm c2}'=(dH_{\rm c2}/dT)_{T=T_{\rm c}} \approx T_c/v_{\rm F}^2$. The orbital limitation at zero temperature is of course proportional to $T_c$ and to $H_{\rm c2}'$, and in a weak coupling scheme, it can be estimated from $H_{\rm orb}(0) = - 0.7 T_{\rm c} H_{\rm c2}'$ \cite{Werthammer1966}. In a strong coupling scheme,  probably more appropriate for the Ce$M$In$_5$ compounds owing to their large specific heat jump ($\Delta C/C$) at $T_c$, $H_{\rm orb}(0)$ is even larger: the general trend of strong coupling regime is that superconductivity is ''reinforced" toward low temperatures. However, in CeRhIn$_5$ as well as in CeCoIn$_5$, $H_{\rm c2}(0)$ is much lower than $- 0.7 T_{\rm c} H_{\rm c2}'$ by at least a factor 2, which points to an additional mechanism controlling the upper critical field.

\begin{table}[b]
\caption{Experimental values of $T_{\rm c}$, $H_{\rm c2} (0)$, the initial slope $H_{\rm c2}'=(dH_{\rm c2}/dT)_{T=T_{\rm c}}$ at $T_{\rm c}$. The orbital limit of the upper critical field is determined using $H_{\rm orb} = - 0.7 T_{\rm c} (dH_{\rm c2}/dT)_{T=T_{\rm c}}$. The superconducting coherence length $\xi_0$  is estimated by the BCS relation 
$\xi_0 = 0.18 \hbar v_{\rm F} / (k_{\rm B} T_{\rm c})$. 
Further parameters for the best fit shown in Fig.~\ref{Hc2_fit} (solid lines) with strong coupling model for the upper critical field are given, $\lambda$ is the coupling parameter, $g$ the gyromagnetic ratio, and $v_{\rm F}$ the Fermi velocity as well as the bare Fermi velocity $v_{\rm F0}$.\\}
\label{table1}
\begin{tabular}{lcccccc}
\hline
$p$ (GPa)  & 1.7 & 2.4 & 2.6 & 2.8 & 3.7 & 4.5  \\
\hline 	
$T_{\rm c}$ (K) & 2.124 & 2.258 & 2.207 & 2.21 & 1.82 & 1.16 \\
$H_{\rm c2}(0)$ (T) & 10.19 & 10.62 & 9.34 & 9.35 & 5.7 & 2.76  \\
$-H_{\rm c2}'$ (T/K) & 17.28 & 22.83 & 20.34 & 19.93 & 13.6 & 7.25 \\
$H_{\rm orb}$ (T)  & 25.7 & 36.1 & 31.4 & 30.8 & 17.3 & 5.9 \\
$\xi_0$ (\AA) &41.2 & 36.9 & 38.5 & 38.43 & 47.35 & 74.75 \\ 
$\lambda$  & 2.04& 2.2 & 2.14 & 2.14 & 1.71 & 1.133 \\
$g$ & 2.3 & 2.45 & 2.75 & 2.75 & 3.2 & 3.2 \\
$v_F$ ($10^3$m/s) & 6.40 & 6.08 & 6.20 & 6.20 & 6.29 & 6.33 \\ 
$v_{F0}$ ($10^3$m/s) & 19.5 &  19.5 &  19.5 & 19.5 & 17 & 13.5 \\ 

\hline
\end{tabular}
\end{table}

\begin{figure}
\begin{center}
\includegraphics[width=0.8\hsize,clip]{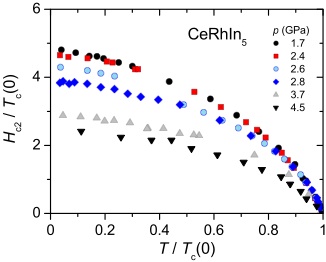}%
 \caption{\label{Hc2_norm}(Color online) Upper critical field of CeRhIn$_5$ (symbols) normalized to the transition temperature $T_{\rm c}$ as function of the reduced temperature $T/T_{\rm c}$ at different pressures.}
\end{center}
 \end{figure}

The other known mechanism is the paramagnetic pair-breaking effect, which originates from Zeeman splitting of single electron energy levels. This so-called Pauli limiting upper critical field can be estimated by $H_{\rm P} = \sqrt{2} \Delta / g \mu_{\rm B}$ \cite{Clogston1962}. Here $\Delta$ is the superconducting energy gap at $T=0$ and $g$ is the gyromagnetic ratio. 
However, the estimation of the $g$ factor is not straightforward: $g=2$ for free electrons, but spin-orbit coupling and Fermi-liquid corrections can lead to strong deviations from this value, as well as the exchange with local moments. An experimental determination from the Pauli susceptibility is also cumbersome in heavy-fermion systems, as the measured susceptibility ($\chi$) mixes Pauli and local contributions, which can also lead to temperature independent terms through Van-Vleck contributions. Moreover, absolute measurements of  $\chi$ are not available under high pressure for CeRhIn$_5$. So in the following, $g$ is considered as a fitting parameter. In Table~\ref{table1} we have summarized the experimentally obtained parameter for CeRhIn$_5$ for different pressures. 

Because in CeRhIn$_5$, $H_{\rm c2}(0)$ seems completely dominated by the paramagnetic limitation, it is nevertheless possible to reveal the evolution of the Pauli limit (equivalently, of the $g$-factor) under pressure without any fit, by using appropriate field and temperature scales to draw $H_{\rm c2}$ in order to suppress the dependence on $T_c $: so Fig.~\ref{Hc2_norm} presents the upper critical field $H_{\rm c2} / T_{\rm c}$ as function of the reduced temperature $T/T_{\rm c}$. The striking feature is that the initial slope at $T_c$ has no strong variation in the pressure range from 1.7~GPa to 2.8~GPa, whereas $H_{\rm c2}(0)$ is systematically depressed with increasing pressure. At first glance, this decrease of $H_{\rm c2}(0)$ with pressure on Fig.~\ref{Hc2_norm} reflects an increase of the gyromagnetic ratio $g$ with pressure. This is true as long as a weak coupling scheme remains valid. But in the Ce$M$In$_5$ compounds, as already mentioned, strong coupling corrections can be large, and one then has to disentangle the pressure evolution of $g$ and of the strong coupling constant $\lambda$, so that  Fig.~\ref{Hc2_norm} could be misleading: a large $\lambda$ leads to a weaker Pauli limitation ($H_{\rm P}(0) \propto \sqrt \lambda$ for $\lambda \gg 1$). However, it will be seen below that quantitatively, even in a strong coupling framework, the $g$ factor is found to increase significantly under pressure. Possible reasons for this behavior will be discussed below. More surprisingly, we would have expected that the strong increase of the effective mass on approaching $p_c$ as observed in dHvA measurements \cite{Shishido2005} would be reflected in a strong enhancement of the initial slope $H_{\rm c2}'$ (remember $H_{\rm c2}' \propto 1/v_{\rm F}^2$).

\begin{figure}
\begin{center}
\includegraphics[width=0.8\hsize,clip]{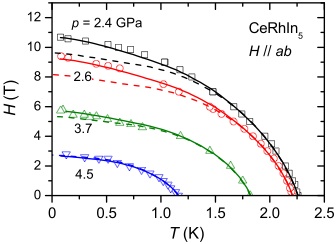}%
 \caption{\label{Hc2_fit}(Color online) Upper critical field of CeRhIn$_5$ at different pressures on a linear scale. Solid lines are fits within a strong coupling model as described in the text. Dashed lines are correspond to fits in a weak coupling approach. }
\end{center}
 \end{figure} 
 
Figure~\ref{Hc2_fit} shows the temperature dependence of $H_{\rm c2}$ of CeRhIn$_5$ for different pressures on a linear scale. It is not possible to fit $H_{\rm c2} (T)$ in a weak coupling approach consistently (see dashed lines in Fig~\ref{Hc2_fit}). If the slope and curvature are well adjusted close to $T_{\rm c}$, then the calculated $H_{\rm c2}$ curve is too low by comparison to the experimental results at low temperatures, notably close to the critical pressure. In fact, this is also true for CeCoIn$_5$, although not publicized as only the low temperature part of the fits is usually shown \cite{Miclea2006}, which can be good at the expense of a bad fit closer to $T_c$. This can be cured by taking strong coupling effects into account (see lines in Fig.~\ref{Hc2_fit}), at the expense of a new parameter, the strong coupling constant $\lambda$. Details of the model are described in refs.~\citen{Thomas1996} and \citen{Glemot1999}. The parameters of the model are (i) the strong coupling parameter $\lambda$, (ii) a characteristic frequency $\Omega$ of the excitations responsible for the pairing ($\lambda$ and $\Omega$ determine completely $T_{\rm c}$, together with the Coulomb repulsion parameter $\mu^\ast$ fixed at the common value $\mu^\ast =0.1$), (iii) the gyromagnetic factor $g$, and (iv) the average Fermi velocity in the plane perpendicular to the applied field. The mass renormalization of the heavy electrons due to the pairing mechanism depends on $\lambda$ according to $m^\star/m_{\rm b} = \lambda +1$; here $m_{\rm b}$ is the band mass of the quasiparticles renormalized by all interactions which do not participate to the pairing potential, and $m^\star$ includes all fluctuations including those contributing to the pairing.

\begin{figure}
\begin{center}
\includegraphics[width=0.8\hsize,clip]{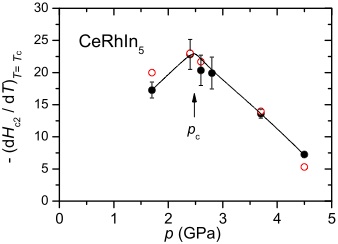}%
 \caption{\label{initial_slope} (Color online) Pressure dependence of the initial slope of the upper critical field $-(dH_{\rm c2}/dT)_{T=T_{\rm c}}$. As criterion for the definition of $T_{\rm c}$ we used the onset of the transition. (Full symbols) are from a linear extrapolation of the data, open symbols from the calculation.}
\end{center}
 \end{figure}

As shown in Fig.~\ref{Hc2_fit} the fit of $H_{\rm c2}$ with $\lambda = 2.2$ (solid lines) at 2.4~GPa is very good up to $p=3.7$~GPa. For these pressures, $\Omega$ and the bare Fermi velocity ($v_{\rm F0}$, not renormalized by the pairing interaction) have been kept fixed. $\lambda (p)$ has been adjusted to reproduce the $T_c (p)$ variation, and it is enough to reproduce also most of the change of the initial slope, whereas the $g$ factor needs to be strongly increased, as anticipated from Fig.~\ref{Hc2_norm}: $H_{\rm c2}'$ depends, like in the weak-coupling scheme, of $T_c$, the physical average Fermi velocity $v_{\rm F} = v_{\rm F0}/(1+\lambda)$, and also slightly on the value of $\lambda$. For $p=4.5$~GPa, the bare Fermi velocity $v_{\rm F0}$ needs to be decreased (by $30\%$) to reproduce the strong slope at fixed $\Omega$, and optimum fit at $p=3.7$~GPa is obtained for a $10\%$ decrease of  $v_{\rm F0}$, although this is almost in the error bars of the experimental points. The parameter of the best fits are also given in Table~\ref{table1}. The starting value of the strong coupling constant of at least $\lambda=2$ is necessary to reproduce the global shape of $H_{\rm c2} (T)$ at the optimum $T_c$, as found also for CeCoIn$_5$. This $\lambda$ value points to an enhancement of the effective mass $m^\star$ due to the pairing potential by $m^\star/m_{\rm b} \approx 3.2$ close to $p_{\rm c}$, which is of same order of magnitute as in CeIn$_3$ or CePd$_2$Si$_2$ (see Tab.~\ref{table3}) \cite{Knebel2001, Sheikin2001}. Figure~\ref{initial_slope} represents the pressure variation of the initial slope which also shows a maximum close to $p_{\rm c}$:  $H'_{\rm c2}$ is highest for maximum $\lambda$, what is natural within the scheme used for the fitting. We found that the variation of the slope with pressure could be reproduced by varying only $\lambda$, keeping $v_{\rm F0}$ constant (at least close to $p_{\rm c}$).

\begin{fulltable}[tb]
\caption{Comparison of parameters of the upper critical field using the strong coupling model for different compounds close to their quantum critical points. (Data for CeCoIn$_5$ are from ref.~\citen{Miclea2006}. In difference to ref.~\citen{Miclea2006} we analyzed the data not in weak coupling model.}
\label{table3}
\begin{fulltabular}{lccccccc}
\\
\hline
		&		& $p\approx p_{\rm c}$~(GPa)& $T_{\rm c} (K)$	&$-H_{\rm c2}'$ (T/K)	& $g$  & $\lambda $ & ref. \\
\hline 	
CeCu$_2$Si$_2$ &$H \parallel a$ & 0 	 & 	0.677	&	23	&	2 & 0.63	& \citen{Sheikin2000}	\\
CePd$_2$Si$_2$ &$H \parallel a$ & 2.67 & 0.395 &	12.7 &  4.6 & 1.5 & \citen{Sheikin2001} \\
			&$H \parallel c$ 	& 	     &		&	16   &  2.35&  1.5 & \citen{Sheikin2001} \\
CeIn$_3$ 	&				&	2.58 & 0.207& 3.2 & 1.4 	& 1.3 & \citen{Knebel2001} \\
CeRhIn$_5$ &$H \parallel a$ & 2.4 & 2.258 & 22.8 & 2.45 & 2.2 & this work \\
CeCoIn$_5$ &$H \parallel a$ & 0  & 2.241 & 30.5 & 2 & 2 & \citen{Miclea2006}, this work \\
          &$H \parallel c$  &    &  & 10.8 & 4.7 & 2 & \citen{Miclea2006}, this work \\
\hline
\end{fulltabular}
\end{fulltable}

\begin{figure}[bt]
\begin{center}
\includegraphics[width=0.8\hsize,clip]{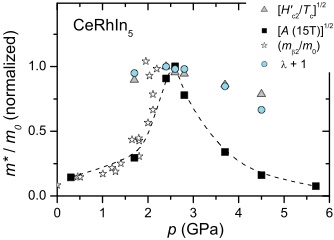}%
\caption{\label{mstar} (Color online) Estimation of the pressure dependence of the effective mass $m^\star$ determined from (i) the initial slope, $m^\star \propto \sqrt{H_{\rm c2}'}/T_{\rm c}$ (triangles), (ii) the $A$ coefficient measured at 15~T, $m^\star \propto \sqrt{A}$ (squares), (iii) the pressure variation of the cyclotron mass of the $\beta_2$ branch from the dHvA experiment from ref.~\citen{Shishido2005} (stars), and the defined from the variation of $\lambda$. All values are normalized to the value at $p_c$, the line is to guide the eyes.    }
\end{center}
 \end{figure}

However, quantitatively, the pressure variation of the effective mass as determined by $\lambda$ (or by $H_{\rm c2}'$ ) derived from resistivity measurements is far weaker than that deduced from the pressure variation of the $A$ coefficient of the resistivity. It is also far weaker than the variation of the cyclotron mass of the $\beta_2$ branch measured directly by de Haas-van Alphen measurements (for $H \parallel c$) \cite{Shishido2005}, which is in good agreement with the estimate from the $A$ coefficient for $p<p_{\rm c}$. This is shown in Fig.~\ref{mstar}. It is difficult to separate the different origins of the mass renormalization of the heavy electrons, but the enhancement $m^\star/m_{\rm b}$ at $p_{\rm c}$ as given from the value of $\lambda$ would correspond also to an increase of the $A$ coefficient by a factor of~9 at the critical pressure (for $\lambda$ ranging between $\approx 2$ at $p_{\rm c}$ down to zero when superconductivity disappears). However, the variation of the $A$ coefficient on the whole pressure range is much larger: $A$ changes more than two orders of magnitude (see Fig.~\ref{Acoeff_P}). Nevertheless, let us note that comparison of the mass enhancement due to strong coupling with the $A$ coefficient for $p>p_{\rm c}$ is very difficult, as a strong variation of the $A$ coefficient may also occur due to a (possible) change of the degeneracy of the Ce.\cite{Miyake1989, Tsujii2005} The relation between $A$ and $\gamma^2$ itself (Kadowaki-Woods rule) changes strongly at $p_{\rm v}$:  entering in the intermediate valence regime the change is by a factor of 15 (see ref.~\citen{Tsujii2005}). In addition, the links between the anisotropy of the $A$ coefficient, the anisotropy of the effective mass $m^\star$ and their respective field dependences are not straightforward.

To summarize, whereas de Haas-van Alphen and transport experiments in the normal state-high field phase, detect a strong enhancement of the effective mass on approaching $p_{\rm c}$, and the de Haas-van Alphen measurements even show that there is an abrupt Fermi surface change at $p_{\rm c}$, the superconducting properties ($H'_{\rm c2}$, $T_{\rm c}$) as determined by the resistivity detect only a smooth evolution crossing $p_{\rm c}$. 
Above $p_{\rm c}^\star$, excellent agreement exist between resistivity and calorimetric measurements in the determination of 
$T_{\rm c}$ and of the slope $H'_{\rm c2}$. 
Below  $p_{\rm c}^\star$ the situation is quite different. $T_{\rm c}(\rho)$ defined by the superconducting anomaly in the resistivity is far higher than $T_{\rm c}(C)$ measured by the ac calorimetry (see Fig.~\ref{phase_diagram}); furthermore, $T_{\rm c}(\rho)$ as well as the slope $H'_{\rm c2}$ derived from the resistivity seems to be the continuation of the high pressure superconducting phase for pressures below $p_{\rm c}^\star$. In the recent ac calorimetry study it is clearly demonstrated that below $p_{\rm c}^\star$ both $T_{\rm c}$ and 
$H'_{\rm c2}$ decrease strongly with decreasing pressure (for $p/p_{\rm c}^\star \sim 0.78$, $T_{\rm c}$ decreases by a factor 4, $H'_{\rm c2}$ by factor of 15, and the jump of the specific heat anomaly by at least a factor of 8).\cite{Park2007b, Park2008b} Thus, as already pointed out previously, the superconducting transition below $p_{\rm c}^\star$ appears quite inhomogeneous as indicated by (i) the large discrepancies in the value of $T_{\rm c}$ determined by different experimental probes, (ii) the strong reduction of the specific heat anomaly at $T_{\rm c}$ and its concomitant large broadening.


 Let us also point out that the fit of the upper critical fields takes into account the appearance of an spatially modulated Fulde-Ferrel-Larkin-Ovchinnikov (FFLO) state. However, this is only for technical reasons, because the calculation of $H_{\rm c2}$ taking orbital and paramagnetic effects into account if it is much simpler if limited to a second order phase transition, which necessarily implies the appearance of such an FFLO state at low temperature for clean systems being so strongly Pauli limited. In our experiment we do not find hints for such a phase transition, but again, new calorimetric experiments to lower temperatures and higher magnetic fields are required. 

We have also delayed the discussion on the evolution of the $g$-factor under pressure: the results in Table~\ref{table1} do show that even with strong coupling corrections, there is a significant increase of the $g$-factor under pressure. At first glance, one possible explanation could be that this is related to the system becoming more isotropic under pressure: if CeRhIn$_5$ under pressure would be like CeCoIn$_5$, with a stronger Pauli limitation along the c axis than along the a axis, then an increase of the isotropy would lead to an increased $g$-factor, and a more isotropic state  is natural under pressure as the Kondo temperature is expected to increase. If it gets of the order of the crystal field splitting, the degeneracy of the ground-state multiplet will increase. As discussed above, there are strong arguments of a possible valence transition~\cite{Miyake2007} in CeRhIn$_5$ at the critical pressure $p_{\rm c} \sim p_{\rm v}$ ((i) linear resistivity, (ii) strong enhancement of the $A$ coefficient at $p_{\rm c}$ and very rapid decrease to very low values for $p>p_{\rm c}$, (iii) maximum of $\rho_0$, (iv) abrupt change of the Fermi surface~\cite{Shishido2005}). So this scenario seems reasonable. However, recent measurements do show that the anisotropy of $H_{\rm c2}$ is reversed in CeRhIn$_5$ in comparison to CeCoIn$_5$.\cite{Ida2008} So at present from the sole CeRhIn$_5$ results, we have no physical interpretation of this increase of the $g$-factor, except that it should be somehow related either to an increase of the ''molecular field" on the conduction electron, or it may be correlated to the suppression of antiferromagnetic correlations or an effect of a multiband system: what is really the passive or active band? Below  we discuss for CeCoIn$_5$ the possible effect a change in the field dependence of the effective mass.

\subsection{Comparison to CeCoIn$_5$}

CeCoIn$_5$ is superconducting at ambient pressure and no magnetism has been observed up to now  at zero magnetic field and on applying high pressure.  However, there are strong evidences that the compound is located close to a quantum critical point due to the observation of strong spin fluctuations and non Fermi liquid behavior such as a linear temperature dependence of the resistivity over a large range of temperature above the superconducting transition.\cite{Kohori2001, Petrovic2001, Nicklas2001, Sidorov2002, Paglione2003}
The application of a magnetic field sufficiently high to suppress superconductivity reveals a specific heat which increases as $-T\ln T$ to low temperatures for both directions, $H \parallel a$ and $H\parallel c$ \cite{Petrovic2001, Ronning2005}. 
The temperature dependence of the resistivity in CeCoIn$_5$ is quite similar to our observation for CeRhIn$_5$ close to $p_{\rm c}$ as shown in Fig.~\ref{nT} with an exponent $n \to 1$ in high magnetic fields. In CeCoIn$_5$ for $H \parallel ab$ and the current in the basal plane at $p = 0$ at e.g.~$H=18$~T, on cooling $n$ starts with a value $n \sim 0.6$ for $T=2$~K, reaches 1.5 at $T \sim 0.2$~K and seems to extrapolate to the Fermi liquid value $n=2$, but de facto decreases towards $n=1$ after passing through a maxima of $n~\sim 1.7$ for $T=90$~mK.\cite{Ronning2005} For CeRhIn$_5$ at high magnetic field of 15~T we observe the same behavior as shown in Fig.~\ref{nT} at 2.4, 2.6, and 3.7~GPa, the temperature where $n$ deviates from the Fermi liquid limit $n = 2$ increases with pressure in good agreement with the idea that this departure from the Fermi liquid regime is just an artifact of the crossing from the conditions $\omega_{\rm c}\tau < 1$ to $\omega_{\rm c}\tau > 1$ on cooling. Thus, it is worthwhile to remember that FL deviations at high magnetic field may not always be linked to the proximity of a quantum critical point, but may also result from the lost of the collision regime (which was characterized by $\omega_{\rm c}\tau \ll 1$) as discussed above. Even for the clean heavy fermion compound UPt$_3$ an increase of $\rho (T)$ has been clearly reported,\cite{Taillefer1988a, Taillefer1988b} as well as in CeCoIn$_5$.\cite{Paglione2003} 

Comparing the $(p, T)$ phase diagram of CeCoIn$_5$ with CeRhIn$_5$ it gets obvious that CeCoIn$_5$ corresponds clearly to the condition $p_{\rm c}^\star < 0$. Applying pressure, suppresses the strong spin fluctuations and tunes the system even away from a quantum critical point.\cite{Yashima2004} The pressure variation of the effective mass  derived from the $A$ coefficient of the resistivity\cite{Sidorov2002}, from the initial slope of the upper critical field\cite{Miclea2006},  the jump of the superconducting specific heat anomaly,\cite{Sparn2002, Knebel2004} or quantum oscillations\cite{Shishido2003} indicates that the critical pressure $p_c$ would even be negative. However, under pressure the superconducting transition temperature is first increasing up to 1.6~GPa which seems to be a characteristic pressure of the systems. Magnetism can be induced in CeCoIn$_5$ by tiny doping of Cd or Hg on the In site indicating the closeness to a magnetic ordered state.\cite{Pham2006, Bauer2008} Furthermore, the behavior under magnetic field at ambient pressure seems to lead to the conclusion that a quantum critical field $H_{\rm QCP}$ occurs in the vicinity of the upper critical field, for both field directions, $H \parallel a$ and $H \parallel c$.\cite{Bianchi2003c, Paglione2003, Ronning2005} The main observation is the strong increase of the $A$ coefficient of the resistivity on approaching $H_{\rm QCP} \approx H_{\rm c2}$ with $A = A_0(H - H_{QCP})^x$ with  $x = 1.37$ for both field directions, $H \parallel c$ and $H \parallel a$.\cite{Paglione2003, Ronning2005} In Fig.~\ref{Afield}(b) we compare the observed field dependence in CeCoIn$_5$ to CeRhIn$_5$ on a reduced scale $A$ $vs$.~$(H-H_{\rm c2})/ H_{\rm c2}$. Clearly no critical behavior is observed in the field dependence close to $p_{\rm c}$ in CeRhIn$_5$.  
Under pressure the field induced quantum critical point in CeCoIn$_5$ moves inside the superconducting dome and  $H_{\rm QCP}$ vanishes close to $p = 1.6$~GPa where $T_{\rm c} (p)$ has a smooth maximum indicating that the critical field $H_{\rm QCP}$ is well separated from the upper critical  field.\cite{Ronning2006} If this field $H_{\rm QCP}$ is associated to the change of the ground state from antiferromagnetism to paramagnetism, $H_{\rm QCP} = H_{\rm M}$ and $p_{\rm c}$ would be at 1.5~GPa. Obviously, no magnetism appears in CeCoIn$_5$ in the normal phase thus $H_{\rm M} < H_{\rm c2}$ for all pressures and all fields. 

By contrast to CeCoIn$_5$ (occurrence of $H_{\rm QCP}$) in CeRhIn$_5$ no strong field dependence of $A$ is observed, see Fig.~\ref{Afield}. That implies that $T_{\rm N} (p)$ may decrease rapidly under pressure in the range $p_{\rm c}^\star < p < p_{\rm c}$ but $H_{\rm M} (T,p)$ continues to be a sharp function of $T$ just below $T_{\rm N}$ (see Fig.\ref{pd17kbar} and ~Fig.~\ref{phase24kbar}) and its low temperature limit is mainly weakly pressure dependent: $H_{\rm M} (T=0) > H_{\rm c2}$ almost up to $p_{\rm c}$. In CeCoIn$_5$, if antiferromagnetism is an underlying mechanism, $H_{\rm M} (T,p) < H_{\rm c2}$ for all pressures and $H_{\rm M} \to 0$ close to $p=1.6$~GPa.
 
\begin{figure}[tb]
\begin{center}
\includegraphics[width=0.8\hsize,clip]{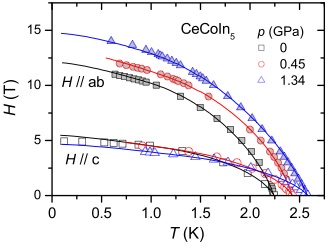}%
 \caption{\label{Miclea_fit_JPB} (Color online) Upper critical field of CeCoIn$_5$ from ref.~\citen{Miclea2006} for $H\parallel ab$ (solid symbols) and $H \parallel c$ (open symbols). Solid lines are fits with the strong coupling model as described in the text. Parameters of the fits are given in Table~\ref{table2}.}
\end{center}
\end{figure}

\begin{table}[tb]
\caption{Experimental values of $T_{\rm c}$, $H_{\rm c2} (0)$, the initial slope $H_{\rm c2}'$ at $T_{\rm c}$, the orbital limit $H_{\rm orb}$ and superconducting coherence length of CeCoIn$_5$ from refs.~\citen{Miclea2006, Nicklas2007b}.  Further parameters of the fits ($\lambda$, $g$ factor, and the Fermi velocity $v_{\rm F}$) shown in Fig.~\ref{Miclea_fit_JPB} (solid lines) with the strong coupling model used for the fitting data of CeRhIn$_5$ are given. We have added experimental (for $H//ab$) and calculated (within an s-wave fit) values for the appearance of the FFLO state. Note that $\xi_0$ has been calculated from $\xi_0 = 0.18 \hbar v_{\rm F} / (k_{\rm B} T_{\rm c})$ and not from $\xi_0 = \sqrt{\Phi_0 / (2 \pi \mu_0 H_{\rm c2}(0))}$ \protect\cite{Miclea2006}, as the strong Pauli limitation governing $H_{\rm c2}(0)$ invalidates the last formula.\\}
\label{table2}
\begin{tabular}{clccc}
\hline
$p$ (GPa) & & 0 & 0.45 & 1.34 \\
\hline 	
				&$T_{\rm c}$ (K) 	&2.241	& 2.425 & 2.58  \\
\hline
$H\parallel ab$ &$H_{\rm c2}(0)$ (T)& 11.6 & 12.5 & 14.3  \\
				&$-H_{\rm c2}'$ (T/K)& 30.5 & 29.4 & 16.4  \\
				&$H_{\rm orb}$ (T)  & 47.9 	& 49.9 	& 29.7 \\
				&$\xi_0$ (\AA) 		& 35 	& 31 	& 36  \\ 
				&$\lambda$  		& 2	& 2.2 	& 2.37  \\
				&$g$ 				& 2	& 2.15	& 2.2 \\
				&$v_{\rm F}$ ($10^3$m/s) 	& 5.8	& 5.5	& 6.8  \\
				&$T_{rm FFLO}-{\rm exp}$ (K) 	& 0.312	& 0.369	& 0.504  \\
				&$T_{\rm FFLO}-{\rm calc}$ (K) 	& 0.9	& 1.1	& 0.88 \\
				
\hline
$H\parallel c$ 	&$H_{\rm c2}(0)$ (T)& 4.9 & 4.7 & 4.2  \\
				&$-H_{\rm c2}'$ (T/K)& 10.8 & 10.2 & 6.5  \\
				&$H_{\rm orb}$ (T)  & 16.9 	& 17.3 	& 11.7 \\
				&$\xi_0$ (\AA) 		& 43 	& 42 	& 48  \\ 
				&$\lambda$  		& 2	& 2.2 	& 2.35  \\
				&$g$ 				& 4.7 	& 5.5	& 7 \\
				&$v_{\rm F}$ ($10^3$m/s) 	& 7.0	& 7.5	& 9.0  \\ 
				&$T_{\rm FFLO}-{\rm calc}$ (K) 	& 1.1	& 1.2	& 1.27  \\

\hline
\end{tabular}
\end{table}

\begin{figure}
\begin{center}
\includegraphics[width=0.8\hsize,clip]{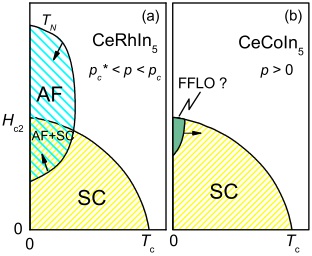}%
 \caption{\label{schematic}(Color online) (a) Schematic $(H, T)$ phase diagram for CeRhIn$_5$ in the pressure range $p_{\rm c}^\star < p < p_{\rm c}$ and (b) for CeCoIn$_5$. The arrows indicate the evolution under pressure. The field induced phase in CeRhIn$_5$ is antiferromagnetic in origin, the low temperature--high magnetic field phase in CeCoIn$_5$ may be a FFLO state. }
\end{center}
 \end{figure}

The upper critical field of CeCoIn$_5$ can be fitted within the same model as presented for CeRhIn$_5$. Figure~\ref{Miclea_fit_JPB} shows the the data of the upper critical field for $H \parallel a$ and $H \parallel c$ from ref.~\citen{Miclea2006} together with fits within the strong coupling model for different pressures. The parameter of the upper critical field are given in Table~\ref{table2}. As mentioned in the previous section, we want to point out that it is not possible to calculate the upper critical field within a weak coupling model in the whole temperature range consistently as opposed to what could be implicitly understood from ref.~\citen{Miclea2006}: the strong coupling effects have to be taken into account. $\lambda$ obtained for CeCoIn$_5$ is comparable to the values in CeRhIn$_5$, CePd$_2$Si$_2$ and CeIn$_3$, see Tables~\ref{table1} and \ref{table3}. The strong anisotropy of the upper critical field is reflected in the anisotropy of the $g$ factor. Furthermore we want to mention that the transition is clearly first order in CeCoIn$_5$, however in CeRhIn$_5$ the situation is less obvious and new thermodynamic measurements are desired.\cite{Bianchi2002, Park2007, Ida2008} This seems to indicate the stronger Pauli limiting in the Co compound.

To understand the $(T, p, H)$ phase digram of CeRhIn$_5$ and CeCoIn$_5$ on the same basis with an antiferromagnetic origin is difficult. As has been discussed above for CeRhIn$_5$, under pressure the field re-entrant AF phase converges to $H_{\rm c2}$ as indicated in Fig.~\ref{schematic}(a). By contrast, the low-temperature high-magnetic field (LTHF) phase in CeCoIn$_5$ exists only inside the superconducting state and expands to higher temperatures under pressure as indicated schematically in Fig.~\ref{schematic}(b).\cite{Miclea2006} 
The ''conventional" explanation for this LTHF phase in CeCoIn$_5$ invokes the achievement of a FFLO state as the Maki parameter $\alpha = \sqrt{2} H_{\rm orb}/H_{\rm P}$ fulfills the required condition $\alpha > 1.8$ \cite{Bianchi2003b, Miclea2006} (for a recent review see ref.~\citen{Matsuda2007} and references therein).
However, the identification of the so-called FFLO state in CeCoIn$_5$ is still under debate. Detailed NMR experiments in CeCoIn$_5$ and on Cd-doped samples have evidenced the existence of static magnetic moments and it can almost be excluded that the high field phase is purely superconducting in origin.~\cite{Mitrovic2006, Young2007, Urbano2007} Looking on the superconducting properties, all conditions for the appearance of a FFLO state seems to be fulfilled in CeRhIn$_5$ too. The paramagnetic limit also exceeds the orbital limit by the same factor than in CeCoIn$_5$, CeRhIn$_5$ is surely in the clean limit, the Fermi surface is highly anisotropic and has a quasi-two dimensional structure~\cite{Hall2001, Shishido2002, Shishido2005}, and the superconducting state has most likely $d_{x^2 - y^2}$ symmetry~\cite{Mito2001}, which is claimed to be favorable to FFLO state.~\cite{Yang1998, Shimahara1994} One can imagine that a FFLO like state could appear above $p_{\rm c}$, however, in our experiment we did not observe any indications of a high field phase for $p>p_{\rm c}$. Remarkably, in CeCoIn$_5$, the FFLO phase extends to higher fields and temperatures under pressure, even if e.g.~the Maki parameter $\alpha$ is reduced,\cite{Miclea2006} while the antiferromagnetic fluctuations are suppressed applying pressure ($H_{\rm QCP} \to 0$): in Table \ref{table2}, it can be seen that the experimental values for the temperature of appearance of the FFLO state are smaller than the theoretical ones, a phenomenon for which many explanations could be put forward. But it is also seen that theory predicts a small decrease of this FFLO sate (for $H \parallel ab)$), whereas a $60\%$ increase is observed between 0 and 1.34 GPa. Moreover, the $g$ factor is found to increase in both directions (particularly along the c axis), whereas the measured dc susceptibility decrease under pressure \cite{Tayama2005}. So, like in CeRhIn$_5$, several points remain unclear as regard the Pauli limitation in the two systems: 

\begin{itemize}
	\item the large increase of the $g$-factor with pressure (for $H \parallel c$ in CeCoIn$_5$, $H \parallel ab$ in CeRhIn$_5$).
	\item the increase of $T_{\rm FFLO}$ under pressure in CeCoIn$_5$, and its absence (for similar conditions) in CeRhIn$_5$.
	\item the opposite anisotropy between $H \parallel c$ and $H \parallel ab$ in the two systems.\cite{Ida2008}
\end{itemize}

At present, there is no satisfying answer to these questions. So it can also be argued that the so-called FFLO phase in CeCoIn$_5$ may have a completely different origin: one could speculate that, if $H_{\rm M} < H_{\rm c2}$, some kind of new magnetic order might appear inside the superconducting phase, taking advantage the opening of the superconducting gap: then  $H_{\rm M}$ would instead stick to $H_{\rm c2}(0)$, and the observed phase would just be a peculiar ''reentrant" magnetic phase. This might be indirectly supported by claims that the antiferromagnetic spin-fluctuations at $p=0$ in CeCoIn$_5$ are unfavorable for the formation of the FFLO state \cite{Tayama2005}. In view of the decrease of the Maki parameter with pressure it will be very informative for the understanding of the interplay of antiferromagnetic spin-fluctuations and the high magnetic field phase, to investigate the $(H$--$T)$ phase diagram of CeCoIn$_5$ to highr pressures. Of course, in the future new ac calorimetry studies are desired to search more thoroughly, even above $p_{\rm c}$, for the existence an FFLO-like state in CeRhIn$_5$.  

The origin of an apparent increase of the $g$ factor derived from the fit of the upper critical field for $H\parallel c$ in CeCoIn$_5$ appears linked to the fact that in CeCoIn$_5$ the effective mass is strongly field dependent for $H > H_{\rm QCP}$. The relative field variation of $m^\star$ below $H_{\rm c2}$ shifts from $H_{\rm c2}(0)$ at $p=0$ to zero at 1.6~GPa. To correct this field change of $m^\star$, an artificial increase of the $g$ factor is necessary in the model which assumes the field invariance of $m^\star$. The importance of a field variation of $m^\star$ is clearly demonstrated for URhGe where the re-entrance  of the superconductivity under magnetic field is strongly connected to the field variation of the effective mass $m^\star$.\cite{Levy2006, Miyake2008}

\section{Conclusions}
In summary, we presented a detailed study of the high pressure phase diagram of CeRhIn$_5$ under high magnetic field. Above $p_{\rm c}^\star$ the phase-diagram determined from these resistivity measurements is in excellent agreement with previous published data from ac-calorimetry. Clear evidence for a quantum critical point at $p_{\rm c} = 2.5$~GPa is given. However, when the superconducting transition temperature $T_{\rm c}$ gets higher than the N\'eel temperature at $p_{\rm c}^\star \approx 2$~GPa the ground state is purely superconducting in zero magnetic field. Under magnetic field an antiferromagnetic state is induced for $p_{\rm c}^\star < p < p_{\rm c}$, which is stable even far above the upper critical field. At $p_{\rm c}$ the antiferromagnetic state collapses. As function of pressure,  a strong enhancement of the inelastic electric scattering term of the resistivity is observed at $p_{\rm c}$. For $p> p_{\rm c}$, no clear Fermi liquid behavior of the resistivity has been recovered. The analysis of the upper critical field allows us to determine the pressure dependence of the strong coupling parameter $\lambda$ which determines the strength of the pairing interaction and of the effective gyromagnetic ratio $g$. The differences to the magnetic field--temperature phase diagram of CeCoIn$_5$ are discussed in detail. 

With the specificity of CeCoIn$_5$, where at $p=0$, $H_{\rm M} < H_{\rm c2}$, the clear event is the field dependence of the effective mass which must play a key role in the temperature and field dependence of the upper critical field $H_{\rm c2}$. It is also suggested that the preservation of a superconducting gap up to $H_{\rm c2}$ may stabilized long range  magnetic order up to $H_{\rm c2}$.

\section*{Acknowledgment} We thank N.~Cherroret for his contributions in the beginning of this work. We acknowledge Profs. R.~Settai and Y.~Onuki for showing us results of the upper critical under high pressure for $H\parallel c$ and allowing us to refer to these results prior publication. 
Financial support has been given by the French ANR within the programs ICENET, ECCE, and NEMSICOM.


\end{document}